\documentclass[10pt,conference,compsocconf]{IEEEtran}
\usepackage[latin9]{inputenc}
\usepackage{color}
\usepackage{array}
\usepackage{verbatim}
\usepackage{multirow}
\usepackage{amsmath}
\usepackage{amsthm}
\usepackage{amssymb}
\usepackage{stmaryrd}
\usepackage{graphicx}
\PassOptionsToPackage{normalem}{ulem}
\usepackage{ulem}

\makeatletter

\providecommand{\tabularnewline}{\\}

  \theoremstyle{definition}
  \newtheorem{defn}{\protect\definitionname}

\usepackage{graphicx}
\usepackage[ruled,vlined,linesnumbered]{algorithm2e}

\usepackage{caption}
\@ifundefined{showcaptionsetup}{}{%
 \PassOptionsToPackage{caption=false}{subfig}}
\usepackage{subfig}
\makeatother

  \providecommand{\definitionname}{Definition}

\begin{document}
\title{A Distributed Collaborative Filtering Algorithm  Using Multiple Data Sources}
\title{\textbf{\Large A Distributed Collaborative Filtering Algorithm  Using Multiple Data Sources}\\[0.2ex]}
\author{\IEEEauthorblockN{~\\[-0.4ex]\large Mohamed Reda Bouadjenek$^{\dag\flat}$, Esther Pacitti$^{\dag}$, Maximilien Servajean$^{\dag}$, Florent Masseglia$^{\dag}$, Amr El Abbadi$^{\ddag}$  \normalsize}
\IEEEauthorblockA{$^{\dag}$INRIA \& LIRMM University of Montpellier France,\\
$^{\flat}$The University of Toronto, Department of Mechanical and Industrial Engineering\\
Email: {mrb@mie.utoronto.ca, \{esther.pacitti, maximilien.servajean\}@lirmm.fr}, florent.masseglia@inria.fr\\
$^{\ddag}$University of California, Santa Barbara, CA, USA, Email: {amr@cs.ucsb.edu}
}}

\newcommand{\subfour}[1]{\vspace*{3mm}{\noindent\bf #1}}  
\newcommand{\subsubfour}[1]{\vspace*{1mm}{\noindent\bf #1}} 
\maketitle
\begin{abstract}
Collaborative Filtering (CF) is one of the most commonly used recommendation
methods. CF consists in predicting whether, or how much, a user will
like (or dislike) an item by leveraging the knowledge of the user's
preferences as well as that of other users. In practice, users interact
and express their opinion on only a small subset of items, which makes
the corresponding user-item rating matrix very sparse. Such data sparsity
yields two main problems for recommender systems: (1) the lack of
data to effectively model users' preferences, and (2) the lack of
data to effectively model item characteristics. However, there are
often many other data sources that are available to a recommender
system provider, which can describe user interests and item characteristics
(e.g., users' social network, tags associated to items, etc.). These
valuable data sources may supply useful information to enhance a recommendation
system in modeling users' preferences and item characteristics more
accurately and thus, hopefully, to make recommenders more precise.
For various reasons, these data sources may be managed by clusters
of different data centers, thus requiring the development of distributed
solutions. In this paper, we propose a new distributed collaborative
filtering algorithm, which exploits and combines multiple and diverse
data sources to improve recommendation quality. Our experimental evaluation
using real datasets shows the effectiveness of our algorithm compared
to state-of-the-art recommendation algorithms. 
\end{abstract}
\IEEEpeerreviewmaketitle 

\section{Introduction}

Nowadays, Internet floods users with useless information. Hence, recommender
systems are useful to supply them with content that may be of interest.
Recommender systems have become a popular research topic over the
past 20 years, to make them more accurate and effective along many
dimensions (social dimension \cite{Ma2008}\cite{Ma2011}\cite{Noel2012},
geographical dimension \cite{Levandoski2012}\cite{Wang2013}, diversification
aspect \cite{Abbar2013}\cite{Ziegler2005}\cite{Zhang2008}, etc.).

Collaborative Filtering (CF) \cite{Resnick1997} is one of the most
commonly used recommendation methods. CF consists in predicting whether,
or how much, a user will like (or dislike) an item by leveraging the
knowledge of the user preferences, as well as that of other users.
In practice, users interact and express their opinions on only a small
subset of items, which makes the corresponding user-item rating matrix
very sparse. Consequently, in a recommender system, this data sparsity
induces two main problems: (1) the lack of data to effectively model
user preferences (new users suffer from the cold-start problem \cite{Sedhain2014}),
and (2) the lack of data to effectively model items characteristics
(new items suffer from the cold-start problem since no user has yet
rated them).

On the other hand, beside this sparse user-item rating matrix, there
are often many other data sources that are available to a recommender
system, which can provide useful information that describe user interests
and item characteristics. Examples of such diverse data sources are
numerous: a user social network, a user's topics of interest, tags
associated to items, etc. These valuable data sources may supply useful
information to enhance a recommendation system in modeling user preferences
and item characteristics more accurately and thus, hopefully, to make
more precise recommendations. Previous research work has demonstrated
the effectiveness of using external data sources for recommender systems
\cite{Ma2008}\cite{Ma2011}\cite{Noel2012}. However, most of the
proposed solutions focus on the use of only one kind of data provided
by an online service (e.g., social network in \cite{Ma2008} or geolocation
information in \cite{Levandoski2012}\cite{Wang2013}). Extending
these solutions into a unified framework that considers multiple and
diverse data sources is itself a challenging research problem.

Furthermore, these diverse data sources are typically managed by clusters
at different data centers, thus requiring the development of new distributed
recommendation algorithms to effectively handle this constantly growing
data. 
In order to make better use of these different data sources, we propose
a new distributed collaborative filtering algorithm, which exploits
and combines multiple and diverse data sources to improve recommendation
quality. To the best of our knowledge, this is the first attempt to
propose such a distributed recommendation algorithm. In summary, the
contributions of this paper are: 
\begin{enumerate}
\item A new recommendation algorithm, based on matrix factorization, which
leverages multiple and diverse data sources. This allows better modeling
user preferences and item characteristics. 
\item A distributed version of this algorithm that mainly computes factorizations
of matrices by exchanging intermediate latent feature matrices in
a coordinated manner. 
\item A thorough comparative analysis with state-of-the-art recommendation
algorithms on different datasets. 
\end{enumerate}
\indent This paper is organized as follows: Section \ref{sec:CaseStudy}
provides two use cases; Section \ref{sec:background} presents the
main concepts used in this paper; Section \ref{sec:recmodel} describes
our multi-source recommendation model; Section \ref{sec:distribution}
gives our distributed multi-source recommendation algorithm; Section
\ref{sec:evaluation} describes our experimental evaluation; Section
\ref{sec:relatedworks} discusses the related work; Finally, Section
\ref{sec:conslusion} concludes and provides future directions.

\section{Use Cases}

\label{sec:CaseStudy}

Let us illustrate our motivation with two use cases, one with internal
data sources, one with external data sources.

\subsection{Diverse internal data sources}

\emph{Consider John, a user who has rated a few movies he saw on a
movie recommender system. In that same recommendation system, John
also expressed his topics of interest regarding movie genres he likes.
He also maintains a list of friends, which he trusts and follows to
get insight on interesting movies. Finally, John has annotated several
movies he saw, with tags to describe their contents.}

In this example, the same recommender system holds many valuable data
sources (topics of interest, friends list, and annotations), which
may be used to accurately model John's preferences and movies' characteristics,
and thus, hopefully to make more precise recommendations. In this
first scenario, we suppose that these diverse data sources are hosted
over different clusters of the same data center of the recommender
system. It is obvious that a centralized recommendation algorithm
induces a massive data transfer, which will cause a bottleneck problem
in the data center. This clearly shows the importance of developing
a distributed recommendation solution.

\subsection{Diverse external data sources}

\begin{figure}
\begin{centering}
\includegraphics[width=6cm]{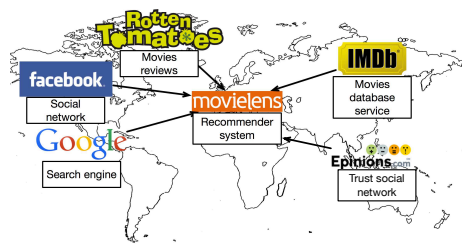} 
\par\end{centering}
\caption{Use case: Movielens.}
\label{fig:UseCase} 
\end{figure}

\emph{Let us now consider that John is a regular user of a movie recommender
system and of many other online services. John uses Google as a search
engine, Facebook to communicate and exchange with his friends, and
maybe other online services such as Epinions social network, IMDb,
which is an online database of information related to films, Movilens,
etc, as illustrated in Figure \ref{fig:UseCase}. }

In this second use case, we believe that by exploiting and combining
all these valuable data sources provided by different online services,
we could make the recommender system more precise. The data sources
are located and distributed over the clusters of different data centers,
which are geographically distributed. In this second use case, we
assume that the recommendation system can identify and link entities
that refer to the same users and items across the different data sources.
We envision that the connection of these online services may be greatly
helped by initiatives like OpenID (http://openid.net/), which promotes
a unified user identity across different services. In addition, we
assume that the online services are willing to help the recommender
system through contracts that can be established. \textcolor{red}{}%

\section{Definitions and Background}

\label{sec:background}

In this section, we introduce the data model we use, and a CF algorithm
based on matrix factorization. Then, we describe the recommendation
problem we study.

\subsection{Data Model}

We use matrices to represent all the data manipulated in our recommendation
algorithm. Matrices are very useful mathematical structures to represent
numbers, and several techniques from matrix theory and linear algebra
can be used to analyze them in various contexts. Hence, we assume
that any data source can be represented using a matrix, whose value
in the $i,j$ position is a correlation that may exist between the
$i^{th}$ and $j^{th}$ elements. We distinguish mainly three different
kinds of data matrices:

\subfour{Users' preferences history:} In a recommender system, there
are two classes of entities, which are referred as users and items.
Users have preferences for certain items, and these preferences are
extracted from the data. The data itself is represented as a matrix
$R$, giving for each user-item pair, a value that represents the
degree of preference of that user for that item. %

\subfour{Users' attributes:} A data source may supply information
on users using two classes of entities, which are referred to users
and attributes. An attribute may refer to any abstract entity that
has a relation with users. We also use matrices to represent such
data, where for each user-attribute pair, a value represents their
correlation (e.g., term frequency-inverse document frequency (tf-idf)
\cite{Baeza-Yates2010}). The way this correlation is computed is
out of the scope of this paper.

\subfour{Items' attributes:} Similarly, a data source may embed information
that describes items using two classes of entities, namely items and
attributes. Here, an attribute refers to any abstract entity that
has a relation with items. Matrices are used to represent these data,
where for each attribute-item pair, a value is associated to represent
their correlation (e.g., tf-idf). The way this correlation is computed
is also beyond the scope of this paper.

Table \ref{tbl:dataattributes} gives examples of attributes that
may describe both users and items, as well as the meaning of the correlations.
It is interesting to notice that these three kinds of matrices are
sparse matrices, meaning that most entries are missing. A missing
value implies that we have no explicit information regarding the corresponding
entry.

\begin{table}[t]
\begin{centering}
\caption{SAMPLE OF ATTRIBUTES.}
{\scriptsize{}\label{tbl:dataattributes}} 
\par\end{centering}
\centering{}%
\begin{tabular}{|c|c|c|}
\hline 
{\scriptsize{}Attribute 1}  & {\scriptsize{}Attribute 2}  & {\scriptsize{}Example of correlation}\tabularnewline
\hline 
\hline 
{\scriptsize{}User}  & {\scriptsize{}User}  & {\scriptsize{}Similarity between the two users}\tabularnewline
\hline 
{\scriptsize{}User}  & {\scriptsize{}Topic}  & {\scriptsize{}Interest of the user in the topic}\tabularnewline
\hline 
\hline 
{\scriptsize{}Item}  & {\scriptsize{}Topic}  & {\scriptsize{}Topic of the items}\tabularnewline
\hline 
{\scriptsize{}Item}  & {\scriptsize{}Item}  & {\scriptsize{}Similarity between two items}\tabularnewline
\hline 
\end{tabular}
\end{table}

\subsection{Matrix Factorization (MF) Models}

Matrix factorization aims at decomposing a user-item rating matrix
$R$ of dimension $I\times J$ containing observed ratings $r_{i,j}$
into a product $R\approx U^{T}V$ of latent feature matrices U and
V of rank K. In this initial MF setting, we designate $U_{i}$ and
$V_{j}$ as the $i^{th}$ and $j^{th}$ columns of $U$ and $V$ such
that $U_{i}^{T}V_{j}$ acts as a measure of similarity between user
$i$ and item $j$ in their respective k-dimensional latent spaces
$U_{i}$ and $V_{j}$.

However, there remains the question of how to learn $U$ and $V$
given that $R$ may be incomplete (i.e., it contains missing entries).
One answer is to define a reconstruction objective error function
over the observed entries, that are to be minimized as a function
of $U$ and $V$, and then use gradient descent to optimize it; formally,
we can optimize the following MF objective \cite{Salakhutdinov2007}:
$\frac{1}{2}{\displaystyle \sum_{i=1}^{I}{\displaystyle \sum_{j=1}^{J}}I_{ij}^{R}(r_{ij}-U_{i}^{T}V_{j})^{2}}$,
where $I_{ij}$ is the indicator function that is equal to 1 if user
$u_{i}$ rated item $v_{j}$ and equal to 0 otherwise. Also, in order
to avoid overfitting, two regularization terms are added to the previous
equation (i.e., $\frac{1}{2}\Vert U\Vert_{F}^{2}$ and $\frac{1}{2}\Vert V\Vert_{F}^{2}$).

\subsection{Problem Definition}

The problem we address in this paper is different from that in traditional
recommender systems, which consider only the user-item rating matrix
$R$. In this paper, we incorporate information coming from multiple
and diverse data matrices to improve recommendation quality. We define
the problem we study in this paper as follows. Given: 
\begin{itemize}
\item a user-item rating matrix $R$; 
\item $N$ data matrices that describe the user preferences $\{S^{U^{1}},\ldots,S^{U^{n}}\}$
distributed over different clusters; 
\item $M$ data matrices that describe the items' characteristics $\{S^{V^{1}},\ldots,S^{V^{m}}\}$
distributed over different clusters; 
\end{itemize}
How to effectively and efficiently predict the missing values of the
user-item matrix $R$ by exploiting and combining these different
data matrices? 

\begin{figure}
\begin{centering}
\includegraphics[width=9.5cm]{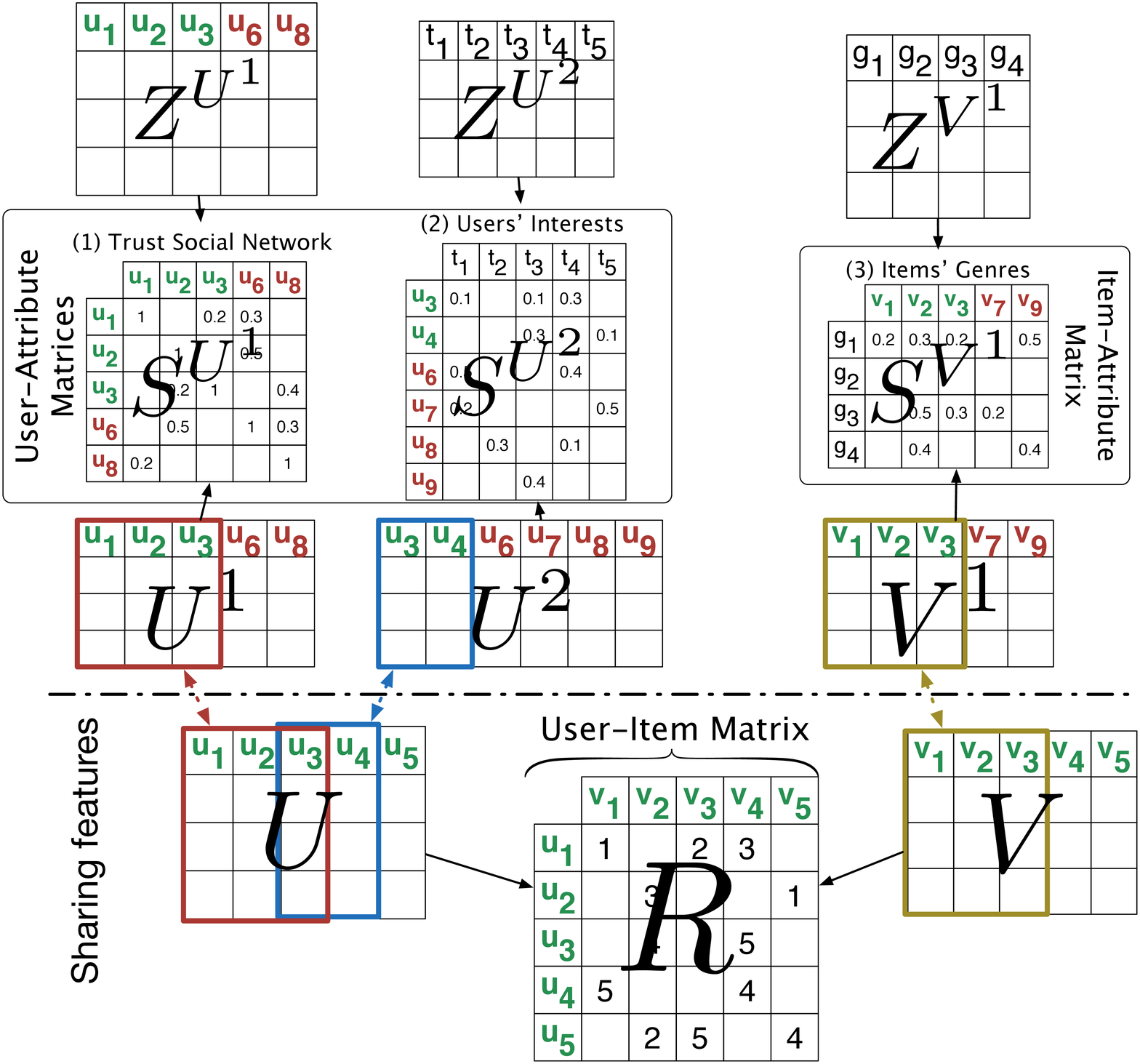} 
\par\end{centering}
\caption{Overview of the recommendation model using toy data. Users and items
in green are users for which we make the recommendation, whereas users
and items in red are used as additional knowledge.}
{\scriptsize{}\label{fig:architecture}}{\scriptsize \par}
\end{figure}

\section{Recommendation Model}

\label{sec:recmodel}

In this section, we first give an overview of our recommendation model
using an example. Then, we introduce the factor analysis method for
our model that uses probabilistic matrix factorization.

\subsection{Recommendation Model Overview}

Let us first consider as an example the user-item rating matrix $R$
of a recommender system (see Figure \ref{fig:architecture}). There
are 5 users (from $u_{1}$ to $u_{5}$) who rated 5 movies (from $v_{1}$
to $v_{5}$) on a 5-point integer scale to express the extent to which
they like each item. Also, as illustrated in Figure \ref{fig:architecture},
the recommender system provider holds three data matrices that provide
information that describe users and items. Note that only part of
the users and items of these data matrices overlap with those of the
user-item rating matrix. 
\begin{enumerate}
\item Matrix (1): provides the social network of $u_{1}$, $u_{2}$, and
$u_{3}$, where each value in the matrix $S^{U^{1}}$ represents the
trustiness between two users%
. 
\item Matrix (2): provides information about the interests of $u_{3}$ and
$u_{4}$, where for each user-topic pair in the matrix $S^{U^{2}}$,
a value is associated, which represents the interest of the user in
this topic. 
\item Matrix (3): provides information about the genre of the movies $v_{1}$,
$v_{2}$, and $v_{3}$ in the matrix $S^{V^{1}}$.
\end{enumerate}
The problem we study in this paper is how to predict the missing values
of the user-item matrix $R$ effectively and efficiently by combining
all these data matrices ($S^{U^{1}}$, $S^{U^{2}}$, and $S^{V^{1}}$).
Motivated by the intuition that more information will help to improve
a recommender system, and inspired by the solution proposed in \cite{Ma2008},
we propose to disseminate the data matrices of the data sources in
the user-item matrix, by factorizing all these matrices simultaneously
and seamlessly as illustrated in Figure \ref{fig:architecture}, such
as: $R\thickapprox U^{T}V$, $S^{U^{1}}\thickapprox U^{1T}Z^{U^{1}}$,
$S^{U^{2}}\thickapprox U^{2T}Z^{U^{2}}$, and $S^{V^{1}}\thickapprox Z^{V^{1}T}V^{1}$,
where the k-dimensional matrices $U$, $U^{1}$, and $U^{2}$ denote
the user latent feature space, such as $U_{1}=U_{1}^{1}$, $U_{2}=U_{2}^{1}$,
$U_{3}=U_{3}^{1}=U_{1}^{2}$, $U_{4}=U_{2}^{2}$ ($U_{1}$, $U_{2}$,
$U_{4}$, and $U_{4}$ refer respectively to the $1^{st}$, $2^{nd}$,
$3^{rd}$, and $4^{th}$ column of the matrix $U$), the matrices
$V$ and $V^{1}$ are the k-dimensional item latent feature space
such as $V_{1}=V_{1}^{1}$, $V_{2}=V_{2}^{1}$, $V_{3}=V_{3}^{1}$,
and $Z^{U^{1}}$, $Z^{U^{2}}$, and $Z^{V^{1}}$, are factor matrices.
In the example given in Figure \ref{fig:architecture}, we use 3 dimensions
to perform the factorizations of the matrices. Once done, we can predict
the missing values in the user-items matrix $R$ using $U^{T}V$.
In the following sections, we present the details of our recommendation
model.

\subsection{User-Item Rating Matrix Factorization}

Suppose that a Gaussian distribution gives the probability of an observed
entry in the User-Item matrix as follows:

\begin{equation}
r_{ij}\sim\mathcal{N}(r_{ij}|U_{i}^{T}V_{j},\sigma_{R}^{2})
\end{equation}

\noindent where $\mathcal{N}(x|\mu,\sigma^{2})$ is the probability
density function of the Gaussian distribution with mean $\mu$ and
variance $\sigma^{2}$. The idea is to give the highest probability
to $r_{ij}\approx U_{i}^{T}V_{j}$ as given by the Gaussian distribution.
Hence, the probability of observing approximately the entries of $R$
given the feature matrices $U$ and $V$ is:

\begin{equation}
p(R|U,V,\sigma_{R}^{2})={\displaystyle \prod_{i=1}^{I}}{\displaystyle \prod_{j=1}^{J}}\left[\mathcal{N}(r_{ij}|U_{i}^{T}V_{j},\sigma_{R}^{2})\right]^{I_{ij}^{R}}
\end{equation}

\noindent where $I_{ij}^{R}$ is the indicator function that is equal
to 1 if a user $i$ rated an item $j$ and equal to 0 otherwise. Similarly,
we place zero-mean spherical Gaussian priors \cite{Dueck2004}\cite{Ma2008}\cite{Salakhutdinov2007}
on user rating and item feature vectors:

\begin{equation}
\begin{array}{c}
p(U|\sigma_{U}^{2})={\displaystyle \prod_{i=1}^{I}}\left[\mathcal{N}(U_{i}|0,\sigma_{U}^{2})\right],p(V|\sigma_{V}^{2})={\displaystyle \prod_{j=1}^{J}}\left[\mathcal{N}(V_{j}|0,\sigma_{V}^{2})\right]\end{array}
\end{equation}

Hence, through a simple Bayesian inference, we have:

\begin{equation}
\begin{array}{c}
p(U,V|R,\sigma_{R}^{2},\sigma_{U}^{2},\sigma_{V}^{2})\propto p(R|U,V,\sigma_{R}^{2})p(U|\sigma_{U}^{2})p(V|\sigma_{V}^{2})\end{array}\label{eq:BaysianInference1}
\end{equation}

\subsection{Matrix factorization for data sources that describe users}

Now let's consider a User-Attribute matrix $S^{U^{n}}$ of $P$ users
and $K$ attributes, which describes users. We define the conditional
distribution over the observed matrix values as:

\begin{equation}
p(S^{U^{n}}|U^{n},Z^{U^{n}},\sigma_{S^{U^{n}}}^{2})={\displaystyle \prod_{p=1}^{P}}{\displaystyle \prod_{k=1}^{K}}\left[\mathcal{N}(s_{pk}^{U^{n}}|U_{p}^{nT}Z_{k}^{U^{n}},\sigma_{S^{U^{n}}}^{2})\right]^{I_{pk}^{S^{U^{n}}}}
\end{equation}

\noindent where $I_{pk}^{S^{U^{n}}}$ is the indicator function that
is equal to 1 if user $p$ has a correlation with attribute $k$ (in
the data matrix $S^{U^{n}}$) and equal to 0 otherwise. Similarly,
we place zero-mean spherical Gaussian priors on feature vectors:

{\tiny{}
\begin{equation}
\begin{array}{cc}
p(U^{n}|\sigma_{U}^{2})={\displaystyle \prod_{p=1}^{P}}\left[\mathcal{N}(U_{p}^{n}|0,\sigma_{U}^{2})\right], & p(Z^{U^{n}}|\sigma_{Z^{U^{n}}}^{2})={\displaystyle \prod_{k=1}^{K}}\left[\mathcal{N}(Z_{k}^{U_{n}}|0,\sigma_{Z^{U^{n}}}^{2})\right]\end{array}
\end{equation}
}{\tiny \par}

Hence, similar to Equation \ref{eq:BaysianInference1}, through a
simple Bayesian inference, we have:

\begin{equation}
\begin{array}{c}
p(U^{n},Z^{U^{n}}|S^{U^{n}},\sigma_{S^{U^{n}}}^{2},\sigma_{U}^{2},\sigma_{Z^{U^{n}}}^{2})\propto\\
\vphantom{{\displaystyle \prod_{i=1}^{I}}}p(S^{U^{n}}|U^{n},Z^{U^{n}},\sigma_{S^{U^{n}}}^{2})p(U^{n}|\sigma_{U}^{2})p(Z^{U^{n}}|\sigma_{Z^{U^{n}}}^{2})
\end{array}
\end{equation}

\subsection{Matrix factorization for data sources that describe items}

Now let's consider an Item-Attribute matrix $S^{V^{m}}$ of $H$ items
and $K$ attributes, which describes items. We also define the conditional
distribution over the observed matrix values as:

{\small{}
\begin{equation}
\begin{array}{l}
p(S^{V^{m}}|V^{m},Z^{V^{m}},\sigma_{S^{V^{m}}}^{2})=\\
{\displaystyle \prod_{h=1}^{H}}{\displaystyle \prod_{k=1}^{K}}\left[\mathcal{N}(s_{hk}^{V^{m}}|V_{h}^{mT}Z_{k}^{V^{m}},\sigma_{S^{V^{m}}}^{2})\right]^{I_{hk}^{S^{V^{m}}}}
\end{array}
\end{equation}
}{\small \par}

\noindent where $I_{hk}^{S^{V^{m}}}$ is the indicator function that
is equal to 1 if an item $h$ is correlated to an attribute $k$ (in
the datasource $S^{V^{m}}$) and equal to 0 otherwise. We also place
zero-mean spherical Gaussian priors on feature vectors:

{\small{}
\begin{equation}
\begin{array}{c}
p(V^{m}|\sigma_{V}^{2})={\displaystyle \prod_{h=1}^{H}}\left[\mathcal{N}(V_{h}^{m}|0,\sigma_{V}^{2})\right]\\
p(Z^{V^{m}}|\sigma_{Z^{V^{m}}}^{2})={\displaystyle \prod_{k=1}^{K}}\left[\mathcal{N}(Z_{k}^{V^{m}}|0,\sigma_{Z^{V^{m}}}^{2})\right]
\end{array}
\end{equation}
}{\small \par}

Hence, through a Bayesian inference, we also have:

\begin{equation}
\begin{array}{c}
p(V^{m},Z^{V^{m}}|S^{V^{m}},\sigma_{S^{V^{m}}}^{2},\sigma_{V}^{2},\sigma_{Z^{V^{m}}}^{2})\propto\\
\vphantom{{\displaystyle \prod_{i=1}^{I}}}p(S^{V^{m}}|V^{m},Z^{V^{m}},\sigma_{S^{V^{m}}}^{2})p(V^{m}|\sigma_{V}^{2})p(Z^{V^{m}}|\sigma_{Z^{V^{m}}}^{2})
\end{array}
\end{equation}

\subsection{Recommendation Model}

\begin{figure}[t]
\begin{centering}
\includegraphics[width=8cm]{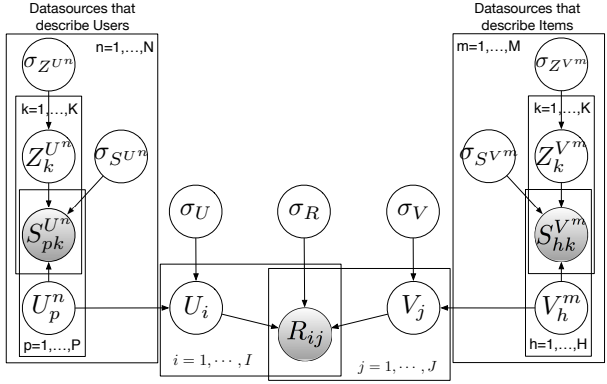} 
\par\end{centering}
\caption{Graphical model for recommendation.}
\label{fig:PGM} 
\end{figure}

Considering $N$ data matrices that describe users, $M$ data matrices
that describe items, and based on the graphical model given in Figure
\ref{fig:PGM}, we model the conditional distribution over the observed
ratings as:{\small{}
\begin{equation}
\begin{array}{l}
p(U,V|R,S^{U^{1}},\ldots,S^{U^{n}},S^{V^{1}},\ldots,S^{V^{m}}\\
\sigma_{Z^{U^{1}}}^{2}\ldots,\sigma_{Z^{U^{n}}}^{2},\sigma_{Z^{V^{1}}}^{2},\ldots,\sigma_{Z^{V^{m}}}^{2})\propto\\
p(R|U,V,\sigma_{R}^{2})p(U|\sigma_{U}^{2})p(V|\sigma_{V}^{2})\\
{\displaystyle \prod_{n=1}^{N}p(S^{U^{n}}|U^{n},Z^{U^{n}},\sigma_{S^{U^{n}}}^{2})p(U^{n}|\sigma_{U}^{2})p(Z^{U^{n}}|\sigma_{Z^{U^{n}}}^{2})}\\
{\displaystyle \prod_{m=1}^{M}p(S^{V^{m}}|V^{m},Z^{V^{m}},\sigma_{S^{V^{m}}}^{2})p(V^{m}|\sigma_{V}^{2})p(Z^{V^{m}}|\sigma_{Z^{V^{m}}}^{2})}
\end{array}
\end{equation}
}{\small \par}

Hence, we can infer the log of the posterior distribution for the
recommendation model as follows:{\scriptsize{} 
\begin{equation}
{\normalcolor \begin{array}{l}
\mathcal{L}(U,U^{1},\ldots,U^{n},V,V^{1},\ldots,V^{m},Z^{U_{1}},\ldots,Z^{U_{n}},Z^{V_{1}},\ldots,Z^{V_{m}})=\\
\left.\frac{1}{2}{\displaystyle \sum_{i=1}^{I}{\displaystyle \sum_{j=1}^{J}}I_{ij}^{R}(r_{ij}-U_{i}^{T}V_{j})^{2}}\right\} \begin{array}{l}
\textrm{Error over the }\\
\textrm{reconstruction of }R
\end{array}\\
\left.+{\displaystyle \sum_{n=1}^{N}\frac{\lambda_{S^{U^{n}}}}{2}{\displaystyle \sum_{p=1}^{P}{\displaystyle \sum_{k=1}^{K}}I_{pk}^{S^{U^{n}}}(s_{pk}^{U^{n}}-U_{p}^{nT}Z_{k}^{U^{n}})^{2}}}\right\} \begin{array}{l}
\textrm{Error over the}\\
\textrm{reconstruction of}\\
\textrm{datasources that }\\
\textrm{describe users}
\end{array}\\
\left.+{\displaystyle \sum_{m=1}^{M}\frac{\lambda_{S^{V^{m}}}}{2}{\displaystyle \sum_{h=1}^{H}{\displaystyle \sum_{k=1}^{K}}I_{hk}^{S^{V^{m}}}(s_{hk}^{V^{m}}-V_{h}^{mT}Z_{k}^{V^{m}})^{2}}}\right\} \begin{array}{l}
\textrm{Error over the}\\
\textrm{reconstruction of}\\
\textrm{datasources that }\\
\textrm{describe items}
\end{array}\\
\left.\begin{array}{l}
+\frac{\lambda_{U}}{2}(\Vert U\Vert_{F}^{2}+{\displaystyle \sum_{n=1}^{N}\Vert U^{n}\Vert_{F}^{2}})\\
+\frac{\lambda_{V}}{2}(\Vert V\Vert_{F}^{2}+{\displaystyle \sum_{m=1}^{M}\Vert V^{m}\Vert_{F}^{2}})\\
+{\displaystyle \sum_{n=1}^{N}\frac{\lambda_{Z^{U^{n}}}}{2}\Vert Z^{U^{n}}\Vert_{F}^{2}}+{\displaystyle \sum_{m=1}^{M}\frac{\lambda_{Z^{U^{m}}}}{2}\Vert Z^{U^{m}}\Vert_{F}^{2}}
\end{array}\right\} \begin{array}{l}
\textrm{Regularization}\\
\textrm{terms}
\end{array}
\end{array}}\label{eq:ObjectiveFunction}
\end{equation}
}{\scriptsize \par}

\noindent where $\Vert.\Vert_{F}^{2}$ denotes the Frobenius norm,
and $\lambda_{*}$ are regularization parameters. A local minimum
of the objective function given by Equation \ref{eq:ObjectiveFunction}
can be found using Gradient Descent (GD) as detailed in Algorithm
\ref{alg:GradientDescent}. We present this algorithm in details in
Appendix \ref{sec:appendix} and a distributed version of this algorithm
in the next section.

\noindent %

\section{Distributed Recommendation}

\label{sec:distribution}

In this section, we first present a distributed version of the Collaborative
Filtering Algorithm \ref{alg:GradientDescent}, which minimizes Equation
\ref{eq:ObjectiveFunction}, and then we carry out a complexity analysis
to show that it can scale to large datasets.

\subsection{Distributed CF Algorithm}

\begin{algorithm}[t]
\scriptsize
\caption{Distributed Collaborative Filtering Algorithm (Master Cluster 1/3)}
\SetAlgoLined
\SetKwData{Left}{left}\SetKwData{This}{this}\SetKwData{Up}{up}
\SetKwFunction{Union}{Union}\SetKwFunction{FindCompress}{FindCompress}
\SetKwInOut{Input}{input}\SetKwInOut{Output}{output}
\Input{The User-Item matrix $R_{ij}$;\\
A learning parameter $\alpha$;\\
Regularization parameters $\lambda_U$, $\lambda_V$;
}
\Output{Feature matrices $U$, $V$;}
\BlankLine

\label{alg:DistGradientDescent1}

Initialize latent feature matrices to small random values.

\label{alg:DistCF1-step1}

\tcc{Minimize $\mathcal{L}$ using gradient descent as follows:}

\While{$\mathcal{L}>\epsilon$ \label{alg:DistCF1-step12}\tcc{$\epsilon$
is a stop criterion}}{

\tcc{Compute local intermediate results of the gradient of $U$ as
follows:}

$\nabla_{U}=\left({\displaystyle I^{R}(U^{T}V-R)V}^{T}\right)^{T}+\lambda_{U}U$

\label{alg:DistCF1-step2}

\ForEach{ user $u_{i}$ }{

\label{alg:DistCF1-step3}

\textbf{\uline{Send}} $U_{i}$ to data sources that share information
about the user $u_{i}$

\label{alg:DistCF1-step4}

}

\ForEach{ $\pi_{n}$ received }{

\label{alg:DistCF1-step5}

\tcc{$\oplus$ is an algebraic operator given in Definition \ref{def:definition1}.}

$\nabla_{U}=\nabla_{U}\oplus\pi_{n}$ 

\label{alg:DistCF1-step6}

}

\tcc{Compute local intermediate results of the gradient of $V$ as
follows:}

$\nabla_{V}=\left({\displaystyle I^{R}(U^{T}V-R)^{T}U^{T}}\right)^{T}+\lambda_{V}V$

\label{alg:DistCF1-step7}

\ForEach{ item $v_{j}$ }{

\textbf{\uline{Send}} $V_{j}$ to data sources that share information
about the item $v_{j}$

}

\ForEach{ $\pi_{m}$ received }{

$\nabla_{V}=\nabla_{V}\oplus\pi_{m}$ 

\label{alg:DistCF1-step8}

}

\tcc{Update global $U$ and $V$ latent features matrices as follows:}

$U=U-\alpha\left(\nabla_{U}\right)$

\label{alg:DistCF1-step9}

$V=V-\alpha\left(\nabla_{V}\right)$

\label{alg:DistCF1-step10}

check $U$ and $V$ for convergence

\label{alg:DistCF1-step11}

}

\end{algorithm}

\begin{algorithm}[t]
\scriptsize
\caption{Distributed Collaborative Filtering Algorithm (User data slave cluster 2/3)}

\label{alg:DistGradientDescent2}

Initialize latent feature matrices $Z^{U^{n}}$ and $U^{n}$ to small
random values.

\SetKwFunction{algo}{algo}\SetKwFunction{proc}{RefineUserFeatures}
\SetKwProg{myproc}{Procedure}{}{}
\myproc{\proc{}}{

\SetAlgoLined
\SetKwData{Left}{left}\SetKwData{This}{this}\SetKwData{Up}{up}
\SetKwFunction{Union}{Union}\SetKwFunction{FindCompress}{FindCompress}
\SetKwInOut{Input}{input}\SetKwInOut{Output}{output}
\Input{A User-Object matrix $S^{U_{n}}$;\\
The common user latent features matrix $U$;\\
A learning parameter $\alpha$;\\
Regularization parameters $\lambda_{S^{U_{n}}}, \lambda_{Z^{U_{n}}}$;
}
\BlankLine

\ForEach{ $U_{i}$ received }{

\label{alg:DistCF2-step1}

Replace the right latent user feature vector $U_{k}^{n}$ with the
received $U_{i}$

\label{alg:DistCF2-step2}

}

\tcc{Compute intermediate result:}

$\pi_{n}=\lambda_{S^{U^{n}}}\left({\displaystyle I^{S^{U^{n}}}(U^{nT}Z^{U^{n}}-S^{U^{n}})Z^{U^{n}}{}^{T}}\right)^{T}$

\label{alg:DistCF2-step3}

Keep in $\pi_{n}$ vectors of users that are shared with the recommender
system

\label{alg:DistCF2-step4}

\textbf{\uline{Send}} $\pi_{n}$ to the recommender system

\label{alg:DistCF2-step5}

\tcc{Compute gradients of $U^{n}$ and $Z$ with respect to $\mathcal{L}$}

{\tiny{}$\nabla_{U^{n}}=\lambda_{S^{U^{n}}}\left({\displaystyle I^{S^{U^{n}}}(U^{nT}Z^{U^{n}}-S^{U^{n}})Z^{U^{n}}{}^{T}}\right)^{T}+\lambda_{S^{U^{n}}}U^{n}$}{\tiny \par}

\label{alg:DistCF2-step6}

{\tiny{}$\nabla_{Z}=\left(\lambda_{S^{U^{n}}}{\displaystyle \left(I^{S^{U^{n}}}(U^{nT}Z^{U^{n}}-S^{U^{n}})\right)^{T}U^{nT}}\right)^{T}+\lambda_{Z^{U^{n}}}Z^{U^{n}}$}{\tiny \par}

\tcc{Update local latent features matrices $U$ and $Z$ as follows:}

$U^{n}=U^{n}-\alpha\left(\nabla_{U^{n}}\right)$

$Z^{U^{n}}=Z^{U^{n}}-\alpha\left(\nabla_{Z}\right)$

\label{alg:DistCF2-step7}

} 

\end{algorithm} 

\begin{algorithm}[t]
\scriptsize
\caption{Distributed Collaborative Filtering Algorithm (Item data slave cluster 3/3)}

\label{alg:DistGradientDescent3}

Initialize latent feature matrices $Z^{V^{m}}$ and $V^{m}$  to small
random values.

\SetKwFunction{algo}{algo}\SetKwFunction{RefineItemFeatures}{proc}
\SetKwProg{myproc}{Procedure}{}{}
\myproc{\proc{}}{
\SetAlgoLined
\SetKwData{Left}{left}\SetKwData{This}{this}\SetKwData{Up}{up}
\SetKwFunction{Union}{Union}\SetKwFunction{FindCompress}{FindCompress}
\SetKwInOut{Input}{input}\SetKwInOut{Output}{output}
\Input{An Object-Item matrix $S^{V_{m}}$;\\
The common user latent features matrix $V$;\\
A learning parameter $\alpha$;\\
Regularization parameters $\lambda_{S^{V_{m}}}, \lambda_{Z^{V_{m}}}$;
}
\BlankLine

\ForEach{ $V_{j}$ received }{

Replace the right latent item feature vector $V_{k}^{m}$ with the
received $V_{j}$

}

\tcc{Compute intermediate result:}

$\pi_{m}=\lambda_{S^{V^{m}}}\left({\displaystyle \left(I^{S^{V^{m}}}(Z^{V^{m}T}V^{m}-S^{V^{m}})\right)^{T}Z^{V^{m}}{}^{T}}\right)^{T}$

Keep in $\pi_{m}$ vectors of items that are shared with the recommender
system

\textbf{\uline{Send}} $\pi_{m}$ to the recommender system

\tcc{Compute gradients of $V$ and $Z$ with respect to $\mathcal{L}$:}

{\tiny{}$\nabla_{V^{m}}=\lambda_{S^{V^{m}}}\left({\displaystyle \left(I^{S^{V_{m}}}(Z^{V^{m}T}V^{m}-S^{V^{m}})\right)^{T}Z^{V^{m}}{}^{T}}\right)^{T}+\lambda_{S^{V^{m}}}V^{m}$}{\tiny \par}

{\tiny{}$\nabla_{Z}=\lambda_{S^{V^{m}}}\left({\displaystyle I^{S^{V^{m}}}(Z^{V^{m}T}V^{m}-S^{V^{m}})V^{T}}\right)^{T}+\lambda_{Z^{U^{m}}}Z$}{\tiny \par}

\tcc{Update local latent features matrices $V$ and $Z$ as follows:}

$V^{m}=V^{m}-\alpha\left(\nabla_{V^{m}}\right)$

$Z^{V^{m}}=Z^{V^{m}}-\alpha\left(\nabla_{Z}\right)$

}

\end{algorithm}

In this section we show how to deploy Algorithm \ref{alg:GradientDescent}
in a distributed setting over different clusters and how to generate
predictions. This distribution is mainly motivated by: (i) the need
to scale up our collaborative filtering algorithm to very large datasets,
i.e., parallelize part of the computation, and (ii) avoid to transfer
the raw data matrices (for mainly privacy concerns). Instead, we only
exchange common latent features.

Based on the observation that several parts of the centralized CF
algorithm (see Algorithm \ref{alg:GradientDescent}) can be executed
in parallel and separately on different clusters, we propose to distribute
it using Algorithms \ref{alg:DistGradientDescent1}, \ref{alg:DistGradientDescent2},
and \ref{alg:DistGradientDescent3}. Algorithm \ref{alg:DistGradientDescent1}
is executed by the master cluster that handle the user-item rating
matrix, whereas each slave cluster that handles data matrices about
users' attributes executes an instance of Algorithm \ref{alg:DistGradientDescent2},
and each slave cluster that hanldes data matrices about items' attributes
executes an instance of Algorithm \ref{alg:DistGradientDescent3}.

Basically, the first step of this distributed algorithm is an initialization
phase, where each cluster (master and slaves) initializes its latent
feature matrices with small random values (lines \ref{alg:DistCF1-step1}
of Algorithms \ref{alg:DistGradientDescent1}, \ref{alg:DistGradientDescent2},
and \ref{alg:DistGradientDescent3}). Next, in line \ref{alg:DistCF1-step2}
of Algorithm \ref{alg:DistGradientDescent1}, the master cluster computes
part of the partial derivative of the objective function given in
Equation \ref{eq:ObjectiveFunction} with respect to $U$ (line \ref{alg:DistCF1-step2}
of Algorithm \ref{alg:DistGradientDescent1} computes a part of line
\ref{alg:GD-Step1} in Algorithm \ref{alg:GradientDescent}). Then,
for each user $u_{i}$, the master cluster sends its latent feature
vector to the other participant slave clusters, which share attributes
about that user (lines \ref{alg:DistCF1-step3} and \ref{alg:DistCF1-step4}
in Algorithm \ref{alg:DistGradientDescent1}). Then, the master cluster
waits for responses of all these participant slave clusters (line
\ref{alg:DistCF1-step5} in Algorithm \ref{alg:DistGradientDescent1}).

Next, each slave cluster that receives users' latent features replaces
the corresponding user latent feature vector $U_{k}^{n}$ with the
user latent feature vector $U_{i}$ received from the master cluster
(lines \ref{alg:DistCF2-step1} and \ref{alg:DistCF2-step2} in Algorithm
\ref{alg:DistGradientDescent2}). Then, the slave cluster computes
$\pi_{n}$, which is part of the partial derivative of the objective
function given in Equation \ref{eq:ObjectiveFunction} with respect
to $U$ (line \ref{alg:DistCF2-step3} in Algorithm \ref{alg:DistGradientDescent2}
computes a part of line \ref{alg:GD-Step1} in Algorithm \ref{alg:GradientDescent}).
Next, the slave cluster keeps in $\pi_{n}$, only vectors of users
that are shared with the master cluster (line \ref{alg:DistCF2-step4}
in Algorithm \ref{alg:DistGradientDescent2}). The slave cluster sends
the remaining feature vectors in $\pi_{n}$ to the master cluster
(line \ref{alg:DistCF2-step5} in Algorithm \ref{alg:DistGradientDescent2}).
Finally, the slave cluster updates its local user and attribute latent
feature matrices $Z^{U^{n}}$ and $U^{n}$ (lines 9-\ref{alg:DistCF2-step7}
in Algorithm \ref{alg:DistGradientDescent2}).

As for the master cluster, each user latent feature matrix $\pi_{n}$
received from a slave cluster is added to $\nabla_{U}$, which is
the partial derivative of the objective function with respect to $U$
(line \ref{alg:DistCF1-step6} in Algorithm \ref{alg:DistGradientDescent1}).
This addition is performed using $\oplus$, an algebraic operator
defined as follows:
\begin{defn}
\label{def:definition1}
\end{defn}
For two matrices $A_{m,n}=(a_{ij})$ and $B_{m,p}=(b_{ij})$, $A\oplus B$
returns the matrix $C_{m,n}$ where:

\[
c_{ij}=\left\{ \begin{array}{ll}
a_{ij}+b_{ij} & \textrm{if }A_{j}=B_{j}\begin{array}{c}
\textrm{(the column refers}\\
\textrm{ to the same user/item)}
\end{array}\\
a_{ij} & \textrm{otherwise}
\end{array}\right.
\]

Once the master cluster has received all the partial derivative of
the objective function with respect to $U$ from all the user participant
sites, it has computed the global derivative of the objective function
given in Equation \ref{eq:ObjectiveFunction} with respect to $U$
. A similar operation is performed for item slave cluster from line
\ref{alg:DistCF1-step7} to line \ref{alg:DistCF1-step8} in Algorithm
\ref{alg:DistGradientDescent1} to compute the global derivative of
the objective function given in Equation \ref{eq:ObjectiveFunction}
with respect to $V$ as given in line \ref{alg:GD-Step2} of Algorithm
\ref{alg:GradientDescent}. Finally, the master cluster updates the
user and item latent feature matrices $U$ and $V$, and evaluates
$\mathcal{L}$ in lines \ref{alg:DistCF1-step9}, \ref{alg:DistCF1-step10},
and \ref{alg:DistCF1-step11} of Algorithm \ref{alg:DistGradientDescent1}
respectively. The convergence of the whole algorithm is checked in
line \ref{alg:DistCF1-step12} of Algorithm \ref{alg:DistGradientDescent1}.
Note that all the involved clusters that hold data matrices on users
and items' attributes execute their respective algorithm in parallel.

\subsection{Complexity Analysis}

The main computation of the GD algorithm evaluates the objective function
$\mathcal{L}$ in Equation \ref{eq:ObjectiveFunction} and its derivatives.
Because of the extreme sparsity of the matrices, the computational
complexity of evaluating the object function $\mathcal{L}$ is $O(\rho_{1}+\ldots+p_{n})$,
where $\rho_{n}$ is the number of nonzero entries in matrix $n$.
The computational complexities for the derivatives are also proportional
to the number of nonzero entries in data matrices. Hence, the total
computational complexity in one iteration of this gradient descent
algorithm is $O(\rho_{1}+\ldots+p_{n})$, which indicates that the
computational time is linear with respect to the number of observations
in the data matrices. This complexity analysis shows that our algorithm
is quite efficient and can scale to very large datasets.

\section{Experimental Evaluation}

\label{sec:evaluation}

In this section, we carry out several experiments to mainly address
the following questions: 
\begin{enumerate}
\item What is the amount of data transferred? 
\item How does the number of user and item sources affect the accuracy of
predictions? 
\item What is the performance comparison on users with different observed
ratings? 
\item Can our algorithm achieve good performance even if users have no observed
ratings? 
\item How does our approach compare with the state-of-the-art collaborative
filtering algorithms? 
\end{enumerate}
In the rest of this section, we introduce our datasets and experimental
methodology, and address these questions (question 1 in Section \ref{subsec:DataTransfer},
question 2 in Section \ref{sec:ImpactNbrSources}, questions 3 and
4 in Section \ref{sec:UserItemDiffRatings}, and question 5 in Section
\ref{sec:Comparaison}).

\subsection{Description of the Datasets}

The first round of our experiment is based on a dataset from Delicious,
described and analyzed in \cite{Wetzker2008}\cite{Bouadjenek2013b}\cite{Bouadjenek2014}
(http://data.dai-labor.de/corpus/delicious/). Delicious is a bookmarking
system, which provides to the user a means to freely annotate Web
pages with tags. Basically, in this scenario we want to recommend
interesting Web pages to users. This dataset contains 425,183 tags,
1,321,039 Web pages, and 318,769 users. The user-item matrix contains
2,265,207 entries (a density of $\simeq0.0005\%$). Each entry of
the user-item matrix represents the degree of which a user interacted
with an item expressed on a $[0,1]$ scale. The dataset contains a
user-tag matrix with 4,598,815 entries, where each entry expresses
the interest of a user in a tag on a $[0,1]$ scale. Lastly, the dataset
contains an item-tags matrix with 4,403,244 entries, where each entry
expresses the coverage of a tag in a Web page on a $[0,1]$ scale.
The user-tag matrix and item-tags are used as user data matrix, and
item data matrix respectively. However, to simulate having many data
matrices that describe both users and items, we have manually and
randomly broken the two previous matrices into 10 matrices in both
columns and rows. These new matrices kept their property of sparsity.
Hence, we end up with a user-item rating matrix, 10 user data matrices
(with approximately 459 000 entries each), and 10 item data matrices
(with approximately 440 000 entries each).

The second round of experiments is based on one of the datasets given
by the HetRec 2011 workshop (http://ir.ii.uam.es/hetrec2011/datasets.html),
and reflect a real use case. This dataset is an extension of the Movielens
dataset, which contains personal ratings, and data coming from other
data sources (mainly IMDb and Rotten Tomatoes). This dataset includes
ratings that range from 1 to 5, including 0.5 steps. This dataset
contains a user-items matrix of 2,113 users, 10,109 movies, and 855,597
ratings (with a density of $\simeq4\%$). This dataset also includes
a user-tag matrix of 9,078 tags describing the users with 21,324 entries
on a $[0,1]$ scale, which is used as a user data matrix. Lastly,
this dataset contains four item data matrices: (1) an item-tag matrix
with 51,794 entries, (2) an item-genre matrix with 20 genres and 20,809
entries, (3) an item-actor matrix with 95,321 actors, and 213,742
entries, and (4) an item-location matrix with 187 locations and 13,566
entries.

\subsection{Methodology and metrics}

We have implemented our distributed collaborative algorithm and integrated
it into Peersim \cite{Montresor2009}, a well-known distributed computing
simulator. %
{} We use two different training data settings (80\% and 60\%) to test
the algorithms. We randomly select part of the ratings from the user-item
rating matrix as the training data (80\% or 60\%) to predict the remaining
ratings (respectively 20\% or 40\%). The random selection was carried
out 5 times independently, and we report the average results. To measure
the prediction quality of the different recommendation methods, we
use the Root Mean Square Error (RMSE), for which a smaller value means
a better performance. We refer to our method Distributed Probabilistic
Matrix Factorization (DPMF). %

\subsection{Data transfer}

\label{subsec:DataTransfer}

Let's consider an example where a recommender system uses a social
network as a source of information to improve its precision. Let's
assume that the social network contains 40 million unique users with
80 billion asymmetric connections (a density of $0.005\%$). It turns
out that if we only consider the connections, the size of the user-user
matrix representing this social network is $80\times10^{9}\times(8\,B+8\,B)\simeq1.16\,TB$
(assuming that we need 8 bytes to encode a double to represent the
strength of the relation between two users, and 8 bytes to encode
a long that represents the key for the entry of the value in the user-user
matrix). Hence, for the execution of the centralized collaborative
filtering algorithm, $1.16\,TB$ of data need to be transferred through
the network. However, if we assume that there are 10\% of common users
between the recommender system and the social network, each iteration
of the DPMF algorithm requires the transfer of $4\times10^{6}\times10\times8\,B\times2\simeq610\,MB$
(assuming that we use 10 dimensions for the factorization, that we
need 8 bytes to encode a double value in a latent user vector, and
a round trip of transfer for the latent vectors in line \ref{alg:DistCF1-step4}
of Algorithm \ref{alg:DistGradientDescent1} and line \ref{alg:DistCF2-step5}
of Algorithm \ref{alg:DistGradientDescent2}). Hence, if the algorithm
requires 100 iterations to converge (roughly the number of iterations
needed in our experiment), the total amount of data transferred is
$59\,GB$, which represents $5\%$ of the data transferred in the
centralized solution. %
{} Finally, the total amount of data transferred depends on the density
of the source, the total number of common attributes, the number of
latent dimensions used for the factorization and the number of iterations
needed for the algorithm to converge. These parameters can make the
DPMF very competitive compared to the centralized solution in terms
of data transfer.

\begin{figure}[t]
\begin{centering}
\includegraphics[width=9cm]{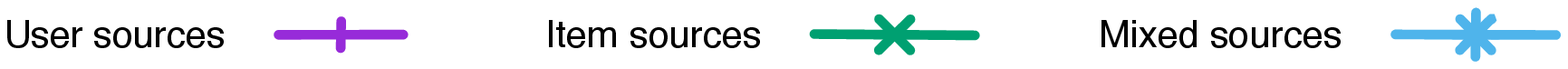} 
\par\end{centering}
\begin{centering}
\subfloat[Training: 80\%.]{\begin{centering}
\includegraphics[width=4.25cm]{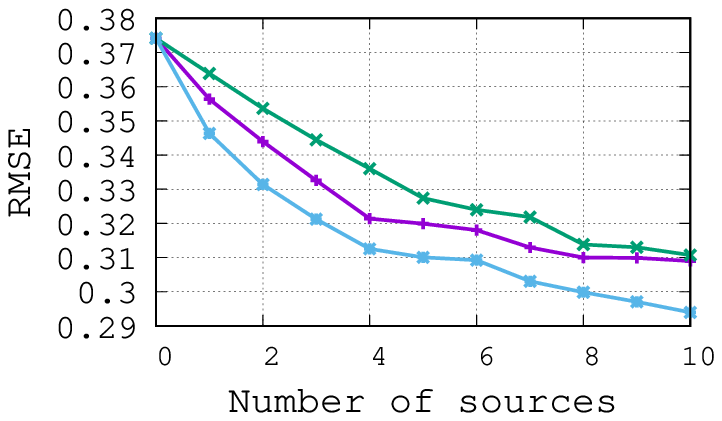}
\par\end{centering}
}\subfloat[Training: 60\%.]{\begin{centering}
\includegraphics[width=4.25cm]{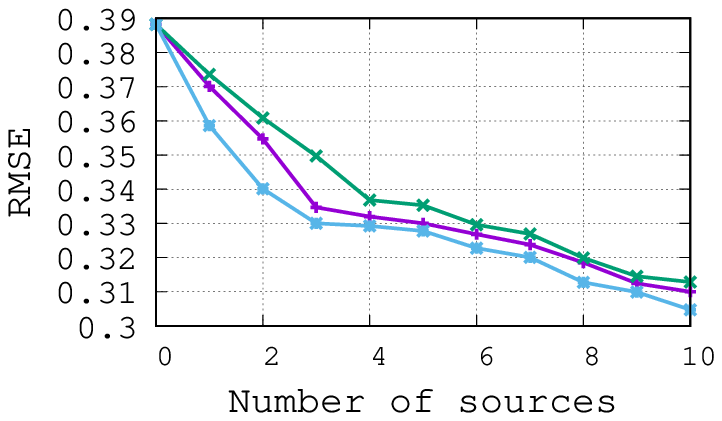}
\par\end{centering}
}
\par\end{centering}
\caption{Results of the impact of the number of sources on the Delicious dataset.}
\label{fig:DeliciousSources} 
\end{figure}

\begin{figure}[t]
\begin{centering}
\subfloat[Training: 80\%.]{\begin{centering}
\includegraphics[width=4.25cm]{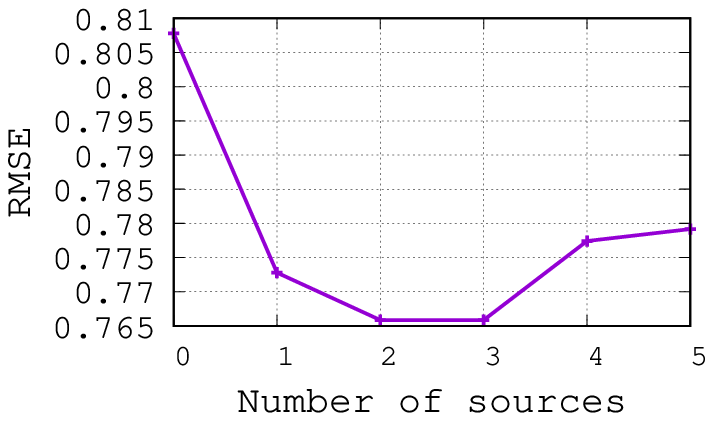}
\par\end{centering}
}\subfloat[Training: 60\%.]{\begin{centering}
\includegraphics[width=4.25cm]{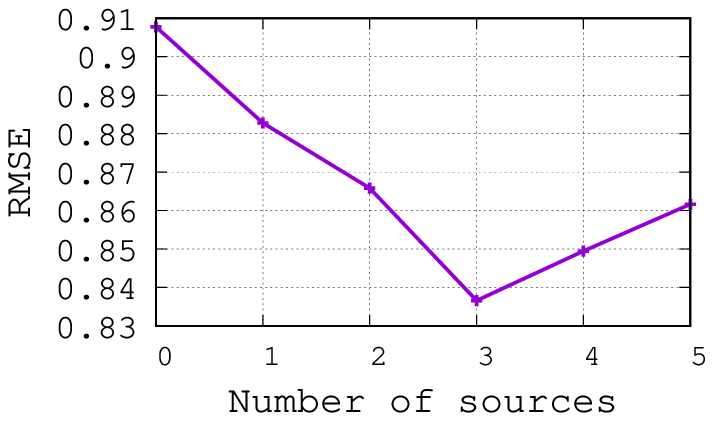}
\par\end{centering}
}
\par\end{centering}
\caption{Results of the impact of the number of sources on the Movielens dataset.
The data matrices are added in the following order: (1) user-tag,
(2) item-tag, (3) item-genre, (4) item-actor, and (5) item-location.}
\label{fig:MovielensSources} 
\end{figure}

\subsection{Impact of the number of sources}

\label{sec:ImpactNbrSources}

Figures \ref{fig:DeliciousSources} and \ref{fig:MovielensSources}
show the results obtained on the two datasets, while varying the number
of sources. Note that source=0 means that we factorize only the user-item
matrix.

In Figure \ref{fig:DeliciousSources}, the green curve represents
the impact of adding item sources only, the red curve the impact of
adding user sources only, and the blue curve the impact of adding
both sources (e.g., 2 sources means we add 2 item and 2 user sources).
First, the results show that adding more sources helps to improve
the performance, confirming our initial intuition. The additional
data sources have certainly contributed to refine users' preferences
and items' characteristics. Second, we observe that sources that describe
users are more helpful than sources that describe items (about 10\%
gain). However, we consider this observation to be specific to this
dataset, and cannot be generalized. Third, we notice that combining
both data sources provides the best performance (blue curve, about
8\% with respect to the use of only user sources). Lastly, the best
gain obtained with respect to the PMF method (source=0) is about 32\%.

Figure \ref{fig:MovielensSources} shows the results obtained on the
Movielens dataset. The obtained results here are quite different than
those obtained on the Delicious dataset. Indeed, we observe that the
data matrices 1, 2 and 3 have a positive impact on the results; however,
data matrices 4 and 5 decrease the performance. This is certainly
due to the fact that the data embedded in these two matrices are not
meaningful to extract and infer items' characteristics. In general,
the best performance is obtained using the three first data matrices,
with a gain of 10\% with respect to PMF (source=0).

\subsection{Performance on users and items with different ratings}

\label{sec:UserItemDiffRatings}

\begin{figure}[t]
\begin{centering}
\subfloat[Different users.]{\begin{centering}
\includegraphics[width=4.25cm]{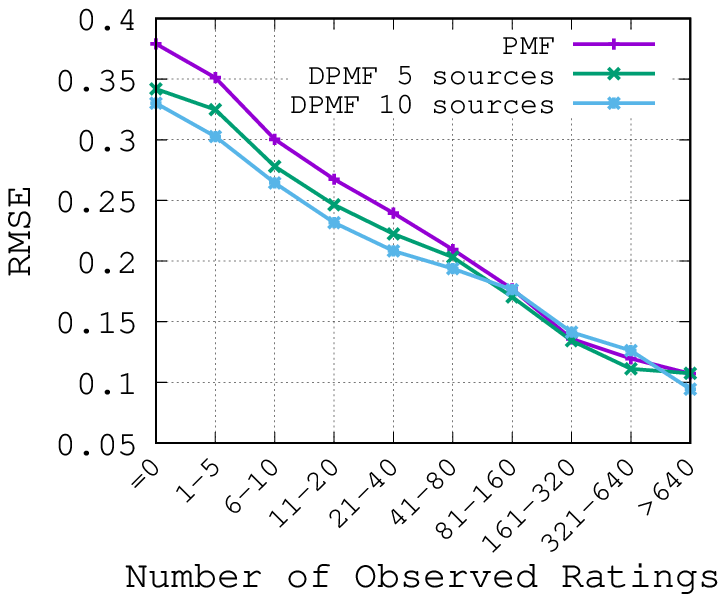}
\par\end{centering}
\label{fig:UsersDiffRatings}}\subfloat[Different items.]{\begin{centering}
\includegraphics[width=4.25cm]{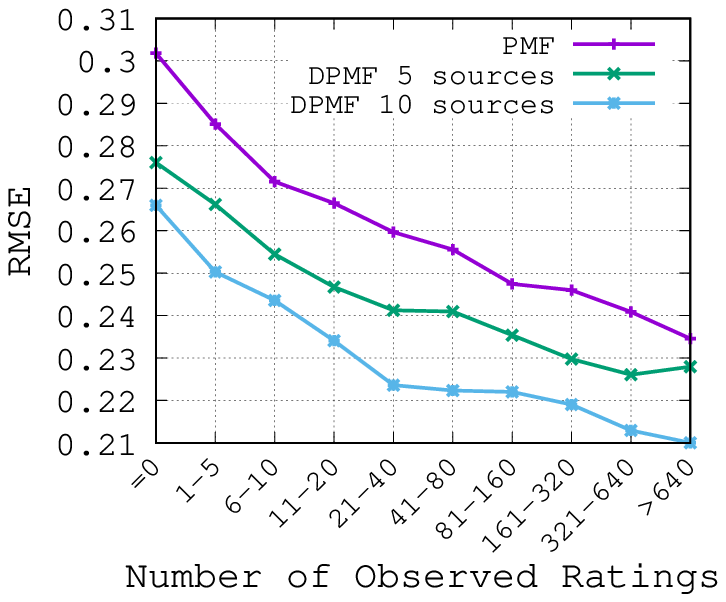}
\par\end{centering}
\label{fig:ItemsDiffRatings}}
\par\end{centering}
\caption{Performance for different ratings on the Delicious dataset.}
\end{figure}

We stated previously that the data sparsity in a recommender system
induces mainly two problems: (1) the lack of data to effectively model
user preferences, and (2) the lack of data to effectively model item
characteristics. Hence, in this section we study the ability of our
method to provide accurate recommendations for users that supply few
ratings, and items that contain few ratings (or no ratings at all).
We show the results for different user ratings in Figure \ref{fig:UsersDiffRatings},
and for different item ratings in Figure \ref{fig:ItemsDiffRatings}
on the Delicious dataset. We group them into 10 classes based on the
number of observed ratings: ``0\textquotedblright , ``1-5\textquotedblright ,
``6-10\textquotedblright , ``11-20\textquotedblright , ``21-40\textquotedblright ,
``41-80\textquotedblright , ``81-160'', ``161-320'', ``321-640'',
and ``\textgreater{}640''. We show the results for different user
ratings in Figure \ref{fig:UsersDiffRatings}, and for different item
ratings in Figure \ref{fig:ItemsDiffRatings} on the Delicious dataset.
We also show the performance of the Probabilistic Matrix Factorization
(PMF) method \cite{Salakhutdinov2007}, and our method using 5 and
10 data matrices. In Figure \ref{fig:UsersDiffRatings}, on the X
axis, users are grouped and ordered with respect to the number of
ratings they have assigned. For example, for users with no ratings
(0 on the X-axis), we got an average of 0.37 for RMSE using the PMF
method. Similarly, in Figure \ref{fig:ItemsDiffRatings}, on the X-axis,
items are grouped and ordered with respect to the number of ratings
they have obtained.

\begin{table*}
\caption{PERFORMANCE COMPARISON USING RMSE. OPTIMAL VALUES OF THE PARAMETERS
ARES USED FOR EACH METHOD (K= 10).}
\label{tbl:Comparaison}
\centering{}{\scriptsize{}}%
\begin{tabular}{|c|c|c|c|c|c|c|c|c|c||c|}
\hline 
{\scriptsize{}Dataset} & {\scriptsize{}Training} & {\scriptsize{}U. Mean} & {\scriptsize{}I. Mean} & {\scriptsize{}NMF} & {\scriptsize{}PMF} & {\scriptsize{}SoRec} & {\scriptsize{}PTBR} & {\scriptsize{}Matchbox} & {\scriptsize{}HeteroMF} & {\scriptsize{}DPMF}\tabularnewline
\hline 
\hline 
\multirow{4}{*}{{\scriptsize{}Delicious}} & \multirow{2}{*}{{\scriptsize{}80\%}} & {\scriptsize{}0.4389} & {\scriptsize{}0.4280} & {\scriptsize{}0.3814} & {\scriptsize{}0.3811} & {\scriptsize{}0.3566} & {\scriptsize{}0.3499} & {\scriptsize{}0.3297} & {\scriptsize{}0.3301} & {\scriptsize{}0.2939}\tabularnewline
 &  & {\scriptsize{}33,03\%} & {\scriptsize{}31,33\%} & {\scriptsize{}22,94\%} & {\scriptsize{}22,88\%} & {\scriptsize{}17.58\%} & {\scriptsize{}16.00\%} & {\scriptsize{}10.85\%} & {\scriptsize{}10.96\%} & {\scriptsize{}Improvement}\tabularnewline
\cline{2-11} 
 & \multirow{2}{*}{{\scriptsize{}60\%}} & {\scriptsize{}0.3965} & {\scriptsize{}0.4087} & {\scriptsize{}0.3779} & {\scriptsize{}0.3911} & {\scriptsize{}0.3681} & {\scriptsize{}0.3599} & {\scriptsize{}0.3387} & {\scriptsize{}0.3434} & {\scriptsize{} 0.3047}\tabularnewline
 &  & {\scriptsize{}23,15\%} & {\scriptsize{}25,44\%} & {\scriptsize{}19,37\%} & {\scriptsize{}22,09\%} & {\scriptsize{}17.22\%} & {\scriptsize{}15.33\%} & {\scriptsize{}10.03\%} & {\scriptsize{}11.26\%} & {\scriptsize{}Improvement}\tabularnewline
\hline 
\multirow{4}{*}{{\scriptsize{}Movielens}} & \multirow{2}{*}{{\scriptsize{}80\%}} & {\scriptsize{}0.8399} & {\scriptsize{}0.8467} & {\scriptsize{}0.7989} & {\scriptsize{}0.8106} & {\scriptsize{}0.774} & {\scriptsize{}0.7801} & \textbf{\scriptsize{}0.7605} & {\scriptsize{}0.7788} & {\scriptsize{}0.7658}\tabularnewline
 &  & {\scriptsize{}8,82\%} & {\scriptsize{}9,55\%} & {\scriptsize{}4,14\%} & {\scriptsize{}5,52\%} & {\scriptsize{}1.05\%} & {\scriptsize{}1.83\%} & \textbf{\scriptsize{}-0.69\%} & {\scriptsize{}1.66\%} & {\scriptsize{}Improvement}\tabularnewline
\cline{2-11} 
 & \multirow{2}{*}{{\scriptsize{}60\%}} & {\scriptsize{}0.9478} & {\scriptsize{}0.9667} & {\scriptsize{}0.9011} & {\scriptsize{}0.9096} & {\scriptsize{}0.882} & {\scriptsize{}0.8912} & {\scriptsize{}0.8399} & \textbf{\scriptsize{}0.8360} & {\scriptsize{}0.8365}\tabularnewline
 &  & {\scriptsize{}11,74\%} & {\scriptsize{}13,46\%} & {\scriptsize{}7,16\%} & {\scriptsize{}8,03\%} & {\scriptsize{}5.15\%} & {\scriptsize{}6.13\%} & {\scriptsize{}0.40\%} & \textbf{\scriptsize{}-0.05\%} & {\scriptsize{}Improvement}\tabularnewline
\hline 
\end{tabular}{\scriptsize \par}
\end{table*}

The results show that our method is more efficient in providing accurate
recommendations compared to the PMF method for both users and items
with few ratings (from 0 to about 100 on the-X axis). Also, the experiments
show that the more we add data matrices, the more the recommendations
are accurate. However, for clarity, we just plot the results obtained
for our method while using 5 and 10 data matrices. Finally, we also
noticed that the performance is better for predicting ratings to items
that contain few ratings, than to users who rated few items. This
is certainly due to the fact that users' preferences change over time,
and thus, the error margin is increased.

\subsection{Performance comparison}

\label{sec:Comparaison}

To demonstrate the performance behavior of our algorithm, we compared
it with eight other state-of-the-art algorithms: \textbf{User Mean:}
uses the mean value of every user; \textbf{Item Mean:} utilizes the
mean value of every item; \textbf{NMF} \cite{Lee1999}; \textbf{PMF}
\cite{Salakhutdinov2007}, \textbf{SoRec} \cite{Ma2008}; \textbf{PTBR}
\cite{Guy2010}; \textbf{MatchBox} \cite{Stern2009}; \textbf{HeteroMF}
\cite{Jamali2013}. The results of the comparison are shown in Table
\ref{tbl:Comparaison}. The optimal parameters of each method are
selected, and we report the final performance on the test set. The
percentages in Table \ref{tbl:Comparaison} are the improvement rates
of our method over the corresponding approaches. 

First, from the results, we see that our method consistently outperforms
almost all the other approaches in all the settings of both datasets.
Our method can almost always generate better predictions than the
state-of-the-art recommendation algorithms. Second, only Matchbox
and HeteroMF slightly outperform our method on the Movielens dataset.
Third, the RMSE values generated by all the methods on the Delicious
dataset are lower than those on Movielens dataset. This is due to
the fact that the rating scale is different between the two datasets.
Fourth, our method outperforms the other methods better on the Delicious
dataset, than on the Movielens dataset (10\% to 33\% on Delicious
and -0.05\% to 11\% on Movielens). This is certainly due to the fact
that: (1) the Movilens dataset contains less data (fewer users and
fewer items), (2) there are less data matrices in the Movielens dataset
to add, and (3) the data matrices of the Delicious dataset are of
higher quality. Lastly, even if we use several data matrices in our
method, using 80\% of training data still provides more accurate predictions
than 60\% of training data. We explain this by the fact that the data
of the user-item matrix are the main resources to train an effective
recommendation model. Clearly, an external source of data cannot replace
the user-item rating matrix, but can be used to enhance it.

\section{Related work}

\label{sec:relatedworks}

\subfour{Enhanced recommendation:} Many researchers have started
exploring social relations to improve recommender systems (including
implicit social information, which can be employed to improve traditional
recommendation methods \cite{Ma2013}), essentially to tackle the
cold-start problem \cite{Lin2013}\cite{Ma2008}\cite{Sedhain2014}.
However, as pointed in \cite{Sedhain2013}, only a small subset of
user interactions and activities are actually useful for social recommendation.

In collaborative filtering based approaches, Liu and Lee \cite{Liu2010}
proposed very simple heuristics to increase recommendation effectiveness
by combining social networks information. Guy et al. \cite{Guy2009}
proposed a ranking function of items based on social relationships.
This ranking function has been further improved in \cite{Guy2010}
to include social content such as related terms to the user. More
recently, Wang et al. \cite{Wang2017} proposed a novel method for
interactive social recommendation, which not only simultaneously explores
user preferences and exploits the effectiveness of personalization
in an interactive way, but also adaptively learns different weights
for different friends. Also, Xiao et al. \cite{Xiao2017} proposed
a novel user preference model for recommender systems that considers
the visibility of both items and social relationships. 

In the context of matrix factorization, following the intuition that
a person's social network will affect her behaviors on the Web, Ma
et al. \cite{Ma2008} propose to factorize both the users' social
network and the rating records matrices. The main idea is to fuse
the user-item matrix with the users' social trust networks by sharing
a common latent low-dimensional user feature matrix. This approach
has been improved in \cite{Ma2009} by taking into account only trusted
friends for recommendation while sharing the user latent dimensional
matrix. Almost a similar approach has been proposed in \cite{Jamali2010}
and \cite{Yang2012} who include in the factorization process, trust
propagation and trust propagation with inferred circles of friends
in social networks respectively. In this same context, other approaches
have been proposed to consider \emph{social regularization terms}
while factorizing the rating matrix. The idea is to handle friends
with dissimilar tastes differently in order to represent the taste
diversity of each user's friends \cite{Ma2011}\cite{Noel2012}. A
number of these methods are reviewed, analyzed and compared in \cite{Yang2014}.

Also, few works consider cross-domain recommendation, where a user's
acquired knowledge in a particular domain could be transferred and
exploited in several other domains, or offering joint, personalized
recommendations of items in multiple domains, e.g., suggesting not
only a particular movie, but also music CDs, books or video games
somehow related with that movie. Based on the type of relations between
the domains, Fernández-Tobías et al. \cite{fernandez2012cross} propose
to categorize cross-domain recommendation as: (i) content based-relations
(common items between domains) \cite{Abel2012}, and (ii) collaborative
filtering-based relations (common users between domain) \cite{Winoto2008}\cite{Berkovsky2007}.
However, almost all these algorithms are not distributed.

\subfour{Distributed recommendation:} Serveral decentralized recommendation
solutions have been proposed mainly from a peer to peer perspective,
basically for collaborative filtering \cite{Kermarrec2012}, search
and recommendation \cite{Draidi2011}. The goal of these solutions
is to decentralize the recommendation process.


Other works have investigated distributed recommendation algorithms
to tackle the problem of scalability. Hence, Liu et al. \cite{Liu2010}
provide a multiplicative-update method. This approach is also applied
to squared loss and to nonnegative matrix factorization with an ``exponential\textquotedblright{}
loss function. Each of these algorithms in essence takes an embarrassingly
parallel matrix factorization algorithm developed previously and directly
distributes it across a MapReduce cluster. Gemulla et al. \cite{Gemulla2011}
provide a novel algorithm to approximately factor large matrices with
millions of rows, millions of columns, and billions of nonzero elements.
The approach depends on a variant of the Stochastic Gradient Descent
(SGD), an iterative stochastic optimization algorithm. Gupta et al.
\cite{Gupta1997} describe scalable parallel algorithms for sparse
matrix factorization, analyze their performance and scalability. Finally,
Yu et al. \cite{Yu2012} uses coordinate descent, a classical optimization
approach, for a parallel scalable implementation of matrix factorization
for recommender system. More recently, Shin et al. \cite{Shin2017}
proposed two distributed tensor factorization methods, CDTF and SALS.
Both methods are scalable with all aspects of data and show a trade-off
between convergence speed and memory requirements.

However, note that almost all the works described above focus mainly
on decentralizing and parallelizing the matrix factorization computation.
To the best of our knowledge, none of the existing distributed solutions
proposes a distributed recommendation approach using diverse data
sources.

\section{Conclusion}

\label{sec:conslusion}

In this paper, we proposed a new distributed collaborative filtering
algorithm, which uses and combines multiple and diverse data matrices
provided by online services to improve recommendation quality. Our
algorithm is based on the factorization of matrices, and the sharing
of common latent features with the recommender system. This algorithm
has been designed for a distributed setting, where the objective was
to avoid sending the raw data, and parallelize the matrix computation.
All the algorithms presented have been evaluated using two different
datasets of Delicious and Movielens. The results show the effectiveness
of our approach. 
Our method consistently outperforms almost all the state-of-the-art
approaches in all the settings of both datasets. Only Matchbox and
HeteroMF slightly outperform our method on the Movielens dataset. 

\bibliographystyle{unsrt}
\bibliography{biblio}

\begin{algorithm*}[t]
\smaller
\caption{Collaborative Filtering Algorithm}
\SetAlgoLined
\SetKwData{Left}{left}\SetKwData{This}{this}\SetKwData{Up}{up}
\SetKwFunction{Union}{Union}\SetKwFunction{FindCompress}{FindCompress}
\SetKwInOut{Input}{input}\SetKwInOut{Output}{output}
\Input{The User-Item matrix $R_{ij}$; Matrices of datasources that describe users: $S_{ik}^{U^{1}},\cdots,S_{ik}^{U^{n}}$; \\ Matrices of datasources that describe items: $S_{ik}^{V^{1}},\cdots,S_{ik}^{V^{m}}$;\\
}
\Output{Feature matrices $U$, $V$}
\BlankLine

\label{alg:GradientDescent}

Initialize feature vectors $U,V,Z^{U^{1}},\ldots,Z^{U^{n}},Z^{V^{1}},\ldots,Z^{V^{m}}$
to small random values.

\tcc{Minimize $\mathcal{L}(U,V,Z^{U^{1}},\ldots,Z^{U^{n}},Z^{V^{1}},\ldots,Z^{V^{m}})$
using gradient descent as follows:}

\While{$\mathcal{L}(U,V,Z^{U^{1}},\ldots,Z^{U^{n}},Z^{V^{1}},\ldots,Z^{V^{m}})>\epsilon$\tcc{$\epsilon$
is a stop criteria}}{

\tcc{Update common Users' Latent Features.}

$U=\overbrace{{\displaystyle U-\alpha\left({\displaystyle I^{R}(U^{T}V-R)V}\oplus\overbrace{\lambda_{S^{U_{1}}}I^{S^{U^{1}}}(U^{1T}Z^{U^{1}}-S^{U^{1}})Z^{U^{1}}}^{\sslash\textrm{Performed by Algorithm \ref{alg:DistGradientDescent2}}}\oplus\cdots\oplus\overbrace{\lambda_{S^{U^{n}}}{\displaystyle I^{S^{U^{n}}}(U^{nT}Z^{U^{n}}-S^{U^{n}})Z^{U^{n}}}}^{\sslash\textrm{Performed by Algorithm \ref{alg:DistGradientDescent2}}}+\lambda_{U}U\right)}}^{\textrm{Performed by Algorithm \ref{alg:DistGradientDescent1}}}$

\label{alg:GD-Step1}

\tcc{Update common Items' Latent Features.}

$V=\overbrace{V-\alpha\left({\displaystyle I^{R}(U^{T}V-R)U}\oplus\overbrace{\lambda_{S^{V^{1}}}{\displaystyle I^{S^{V^{1}}}(Z^{V^{1}T}V^{m}-S^{V^{1}})Z^{V^{1}}}}^{\sslash\textrm{Performed by Algorithm \ref{alg:DistGradientDescent3}}}\oplus\cdots\oplus\overbrace{\lambda_{S^{V^{m}}}{\displaystyle I^{S^{V^{m}}}(Z^{V^{m}T}V^{m}-S^{V^{m}})Z^{V^{m}}}}^{\sslash\textrm{Performed by Algorithm \ref{alg:DistGradientDescent3}}}+\lambda_{V}V\right)}^{\textrm{Performed by Algorithm \ref{alg:DistGradientDescent1}}}$

\label{alg:GD-Step2}

$\left.\begin{array}{l}
\sslash\left\{ U^{1}=U^{1}-\alpha\left(\lambda_{S^{U_{1}}}{\displaystyle I^{S^{U^{1}}}(U^{1T}Z^{U^{1}}-S^{U^{1}})Z^{U^{1}}}+\lambda_{S^{U^{1}}}U^{1}\right)\right.\\
\vdots\\
\sslash\left\{ U^{n}=U^{n}-\alpha\left(\lambda_{S^{U^{n}}}{\displaystyle I^{S^{U^{n}}}(U^{nT}Z^{U^{n}}-S^{U^{n}})Z^{U^{n}}}+\lambda_{S^{U^{n}}}U^{n}\right)\right.
\end{array}\right\} $\tcc{%
\begin{tabular}{c}
Update Non common Users' Latent \tabularnewline
Features performed by Algorithm \ref{alg:DistGradientDescent2}\tabularnewline
\end{tabular}}

$\left.\begin{array}{l}
\sslash\left\{ V^{1}=V^{1}-\alpha\left(\lambda_{S^{V_{1}}}{\displaystyle I^{S^{V_{1}}}(Z^{V_{1}T}V^{1}-S^{V_{1}})Z^{V_{1}}}+\lambda_{S^{V_{1}}}V^{1}\right)\right.\\
\vdots\\
\sslash\left\{ V^{m}=V^{m}-\alpha\left(\lambda_{S^{V^{m}}}{\displaystyle I^{S^{V^{m}}}(Z^{V^{m}T}V^{m}-S^{V^{m}})Z^{V^{m}}}+\lambda_{S^{V^{m}}}V^{m}\right)\right.
\end{array}\right\} $\tcc{%
\begin{tabular}{c}
Update Non common Items' Latent\tabularnewline
Features performed by Algorithm \ref{alg:DistGradientDescent3}\tabularnewline
\end{tabular}}

$\left.\begin{array}{l}
\sslash\left\{ Z^{U^{1}}=Z^{U^{1}}-\alpha\left(\lambda_{S^{U^{1}}}{\displaystyle I^{S^{U^{1}}}(U^{1T}Z^{U^{1}}-S^{U^{1}})U^{1T}}+\lambda_{Z^{U^{1}}}Z\right)\right.\\
\vdots\\
\sslash\left\{ Z^{U^{n}}=Z^{U^{n}}-\alpha\left(\lambda_{S^{U^{n}}}I^{S^{U^{n}}}(U^{nT}Z^{U^{n}}-S^{U^{n}})U^{nT}+\lambda_{Z^{U^{n}}}Z\right)\right.
\end{array}\right\} $\tcc{%
\begin{tabular}{c}
Update Latent Features of Datasources\tabularnewline
that describe users performed by Algorithm \ref{alg:DistGradientDescent2}\tabularnewline
\end{tabular}}

$\left.\begin{array}{l}
\sslash\left\{ Z^{V^{1}}=Z^{V^{1}}-\alpha\left(\lambda_{S^{V^{1}}}{\displaystyle I^{S^{V^{1}}}(Z^{V^{1}T}V^{1}-S^{V^{1}})V^{1}}+\lambda_{Z^{U^{1}}}Z\right)\right.\\
\vdots\\
\sslash\left\{ Z^{V^{m}}=Z^{V^{m}}-\alpha\left(\lambda_{S^{V^{m}}}{\displaystyle I^{S^{V^{m}}}(Z^{V^{m}T}V^{m}-S^{V^{m}})V^{m}}+\lambda_{Z^{U^{m}}}Z\right)\right.
\end{array}\right\} $\tcc{%
\begin{tabular}{c}
Update Latent Features of Datasources\tabularnewline
that describe items performed by Algorithm \ref{alg:DistGradientDescent3}\tabularnewline
\end{tabular}} 

} 

\tcc{$\sslash$ are parts that may be computed separately and in
parallel with no dependency.}

\end{algorithm*}

\appendix 

\section{GRADIENT-BASED OPTIMIZATION}

\label{sec:appendix} We seek to optimize the sum of the objective
function in Equation \ref{eq:ObjectiveFunction} and we use gradient
descent for this purpose in Algorithm \ref{alg:GradientDescent}.
$\sslash$ are parts that may be computed separately and in parallel
with no dependency. These parts are computed by Algorithms \ref{alg:DistGradientDescent1},
\ref{alg:DistGradientDescent2}, and \ref{alg:DistGradientDescent3}.
$\oplus$ is the algebraic operator given in Definition \ref{def:definition1}.
\end{document}